\documentclass[12pt,preprint]{aastex}

\newcommand{\ms}{$M_{\odot}$}
\newcommand{\msb}{$M_{\odot}$~}

\newcommand{\ctb}{$^{13}$C~}

\begin{document}

\title{Can Extra Mixing in RGB and AGB Stars\\
 Be Attributed to Magnetic Mechanisms?}

\author{Maurizio Busso \altaffilmark{1}, Gerald J. Wasserburg\altaffilmark{2}, Kenneth M.
Nollett \altaffilmark{3}, Andrea Calandra\altaffilmark{1}}

\altaffiltext{1}{Department of Physics, University of Perugia, via
Pascoli, Perugia, Italy, 06123; busso@fisica.unipg.it}
\altaffiltext{2}{The Lunatic Asylum, Division of Geology \&
Planetary Sciences, California Institute of Technology, Pasadena, Ca
91125, USA; gjw@gps.caltech.edu} \altaffiltext{3}{Physics Division,
Argonne National Laboratory, Argonne, IL 60439-4843, USA;
nollett@anl.gov}

\begin{abstract}
%From the record of isotopic abundances of light and intermediate elements
%in RGB and AGB stars
It is known that there must be some weak form of transport (called
cool bottom processing, or CBP) acting in low mass RGB and AGB
stars, adding nuclei, newly produced near the hydrogen-burning
shell, to the convective envelope. We assume that this extra-mixing
originates in a stellar dynamo operated by the differential rotation
below the envelope, maintaining toroidal magnetic fields near the
hydrogen-burning shell. We use a phenomenological approach to the
buoyancy of magnetic flux tubes, assuming that they induce matter
circulation as needed by CBP models. This establishes requirements
on the fields necessary to transport material from zones where some
nuclear burning takes place, through the radiative layer, and into
the convective envelope. Magnetic field strengths are determined by
the transport rates needed by CBP for the model stellar structure of
a star of initially 1.5 \ms, in both the AGB and RGB phases. The
field required for the AGB star in the processing zone is $B_0 \sim
5\times10^6$ G; at the base of the convective envelope this yields
an intensity $B_{E} \lesssim 10^4$ G. For the RGB case, $B_0 \sim 5
\times 10^4 - 4\times 10^5$ G, and the corresponding $B_{E}$~ are
$\sim 450 - 3500$ G. These results are consistent with existing
observations on AGB stars. They also hint at the basis for high
field sources in some planetary nebulae and the very large fields
found in some white dwarfs. It is concluded that transport by
magnetic buoyancy should be considered as a possible mechanism for
extra mixing through the radiative zone, as is required by both
stellar observations and the extensive isotopic data on
circumstellar condensates found in meteorites.
\end{abstract}


\section{Introduction}

Many aspects of the physics and evolution of low mass stars are not
adequately treated in available stellar models. One of these problems
is revealed by the photospheric composition of evolved red giants and
by presolar dust grains, preserved in meteorites, which are of
circumstellar origin. In particular, the isotopic admixture of CNO and
the abundances of a few other species (e.g. $^7$Li) in these
environments cannot be reproduced using ``{\it first principles}''
evolutionary codes (see e.g. Pilachowsky et al. 1993; Grundahl et
al. 2002; Wasserburg et al. 2006; Charbonnel \& Do Nascimento
1998). This evidence suggests that mixing mechanisms down to depths
just above the H burning shell must be active during the Red Giant
Branch (RGB) and during the Asymptotic Giant Branch (AGB) phases
(Gilroy \& Brown 1991, Boothroyd, Sackmann, \& Wasserburg 1994). While
abundance problems requiring prolonged mixing already exist on the
Main Sequence, we are here specifically interested in the chemical
anomalies found to occur after the {\it luminosity bump} appears on
the RGB and then continuing to occur through the AGB phase. Reviews of
these phenomena can be found e.g in Kraft (1994) and Charbonnel (2004).

It is now generally agreed that additional transport mechanisms with
low mass transfer rates somehow link the convective envelope to
stable radiative regions where some substantial nuclear processing
occurs. This requires slow movements of mass to take place inside
what is considered as the standard ``stable'' radiative zone above
the H shell.  Parameterized, ad-hoc calculations have been
presented, explaining the above mentioned chemical and isotopic
abundance peculiarities by assuming that material from the
convective envelope is brought down, exposed to partial H burning
(in the so-called {\it Cool Bottom Processes}, or CBP) and then
returned through the radiative zone to the convective envelope by
some form of weak circulation or diffusion (see Wasserburg et al.
1995; Charbonnel \& Do Nascimento 1998; Nollett et al. 2003; Herwig
2005).  The same parameterized scheme devised to account for the
above observational data, also provides clear explanations for the
measurements of high $^{26}$Al/$^{27}$Al ratios, and of distinctive
$^{18}$O/$^{16}$O, $^{17}$O/$^{16}$O and $^{13}$C/$^{12}$C ratios,
in circumstellar grains found in meteorites that are of AGB origin
(Choi et al. 1998; Amari et al. 2001; Nittler 2005; Wasserburg et
al. 2006). The oxygen data, in particular, require extensive
destruction of $^{18}$O and enhanced production of $^{17}$O (cf.
Alexander \& Nittler 1999; Nollett et al. 2003; Clayton \& Nittler
2004).  Figures 6, 7, and 16 of Nollett et al. (2003) show clear
evidence for the requirements of extra mixing in circumstellar dust
grains from AGB stars preserved in meteorites; further evidence from
stellar observations is also discussed there.  It can be seen from
those calculations that to explain the data, processing temperatures
($T_P$) must be close to that of the H-burning shell, and that the
rates of mass transfer must be around $\dot{M}\sim
10^{-6}M_\odot$/yr (see Section 3 below).

Rotation, through shear instabilities and/or meridional circulation,
has been suggested as the physical cause of slow-deep mixing (Zahn
1992; Denissenkov \& Weiss 1996), though alternative mechanisms
have been presented (Denissenkov and Tout 2003; Charbonnel \& Zahn 2007).
Despite formal differences, most models assume that the chemical
(and angular momentum) transport has a diffusive nature, with the
diffusion coefficient left as a free parameter to match the
observations (see e.g. Denissenkov et al. 1998). A similar approach
is commonly used in modeling massive, radiatively stratified stars
(see e.g. Maeder \& Meynet 2004a, b, and references therein).

For low mass red giants the idea of a purely rotation-induced mixing
has recently met difficulties, as the star counter-reacts rapidly in
such a way that any mixing mechanism is of short duration, and any
isotopic change is quenched (Palacios et al. 2006). Similar problems
were found by Siess et al. (2004), while looking for a
rotationally-induced formation of the \ctb reservoir (often called
the \ctb {\it pocket}: cf. Busso et al. 1999), which is necessary to
produce the $s$-elements in He-burning layers. In contrast, the
proposal by Boothroyd et al. (1994) and Wasserburg et al. (1995) of
``Cool Bottom Processing'' does not specify a driving mechanism, but
uses only the path integral of the nuclear reactions and the bulk
mass transport rates. This gives, for the appropriate time scale,
the resulting chemical and isotopic evolution of the convective
envelope for a very wide set of scenarios and provides explicit
values for the temperature required to give the appropriate nuclear
processing, the corresponding pressure (from the stellar model) and
the rates of mass transfer through the processing region.  This
approach and the stellar models with an adjustable diffusion
parameter are essentially equivalent in all results, except that the
time scale for transit of processed material to the convective
envelope may be much faster for the CBP model.

In a recent paper, Eggleton et al. (2007) reported that mixing might
occur due to Rayleigh-Taylor instabilities originating by the
inversion in the molecular weight ($\mu$) induced by $^3$He burning
above the H shell. Calculations were done by making a local 3D
hydrodynamical model of the thin layers where $\mu$ decreases, taking
the inputs from a 1D stellar code at the RGB phase.  The induced
mixing appears sufficient to destroy $^3$He at the surface. Charbonnel
\& Zahn (2007) interpreted this mixing event as due to the double
diffusive mechanism called the {\it thermohaline} instability, early
described by Ulrich (1972). They presented a model in which this
diffusive process efficiently diminishes $^3$He in the envelope and
gives the required shifts in $^{12}$C/$^{13}$C. The extent to which
these recent suggestions are pertinent to the isotopic and chemical
shifts on the AGB is still unknown. In our report we will focus on the
transport mechanism using the CBP model circulation rates and
temperatures and use recent standard stellar structure models to
explore the problem.

Because there is no first-principles transport model, we here seek to
explore, in a preliminary way, whether the circulation can be
accounted for by the buoyancy of magnetic flux tubes that might form
in the neighborhood of the nuclear processing region and transport
matter upward to the base of the convective envelope. It is likely
that differential rotation in such stars can maintain a magnetic
dynamo; this problem is very complex, and no fully self consistent
treatment has yet been found.  Certainly, magnetic fields arise by
some mechanism.

It has been long established that magnetic flux tubes in a star would
provide a buoyant transport mechanism (Parker 1974; 1994) and also
that the oscillations of such tubes (Alfv\'en waves) would evolve into
a number of instabilities that could add to the general buoyancy
(Tayler 1973; Parker 1974; Spruit 1999, 2002).  These phenomena are
usually discussed in the context of determining whether they can
generate and maintain the dynamo, but this is a complex and unresolved
problem that we do not attempt to address.

What we seek to test here is whether magnetic buoyancy phenomena,
considered in a zero-order, general treatment, can account for the
mass circulation rates inferred from parameterized CBP
models. Specifically, we want to establish if they can do so in a
model containing toroidal fields with reasonable values of the
controlling parameters (magnetic field intensities, buoyancy
velocities, fraction of the total radiatively-stratified mass that
must be involved in the mixing mechanism). Such a simple,
order-of-magnitude approach is necessary in order to test whether the
development of a detailed MHD model would be of merit.

We shall try to find general guidance on these issues from
consideration of solar analogies, as magnetic fields inside AGB
stars are essentially unknown. The atmospheres of evolved red giants
rotate so slowly that X-ray observations  from ROSAT, XMM and
CHANDRA now exclude the systematic formation of AGB coronae (see
e.g. Ayres et al 1991; H\"unsch et al. 1996). Even in the bright
prototype Mira system $o$ Cet, the observed X-ray fluxes ($\sim$
2$\times$10$^{29}$ erg/sec) cannot be attributed firmly to the AGB
star (Soker \& Kastner 2003). However, small magnetic fields (few
Gauss) have now been shown to be present in the photospheres of
evolved red giant stars, with the expected toroidal geometry induced
by a dynamo. This was inferred from inspection of circumstellar
emission, in particular from SiO masers formed near the surface
(Herpin et al. 2006).
%The structure of red giant stars is quite
%different from the Sun. They have extended radiative zones (up to
%$\sim$ 1 $R_{\odot}$) overlain by huge convective envelopes (up to
%hundreds of $R_{\odot}$). For a comparison, typical dimensions and
%physical properties of the radiative sub-envelope layers, in a 1.5
%\msb star of half solar metallicity, are shown in Table 1, both
%during the RGB phase (panel b) and during the thermally-pulsing AGB
%stage (panel a). The Table also presents the pertinent solar values
%in (panel c).

\section{Tachoclines  in the Sun and in evolved stars}

\subsection{The solar scenario}

It has been known for more than 80 years (von Zeipel 1924; Eddington
1925) that stars in which energy is transported radiatively cannot
rotate as rigid bodies.  The centrifugal force from rotation deforms
the equipotential surfaces into ellipsoids while radiation retains
spherical symmetry, so radiative and hydrostatic equilbria cannot be
maintained simultaneously.  This conflict drives meridional
circulations known as the Eddington-Sweet effect (Eddington 1925;
Sweet 1950).  Meridional circulation in a radiatively stratified
star implies a transport of angular momentum that drives a state of
differential rotation.

According to studies performed by Spruit (1999, 2002) on the basis
of previous work by Tayler (1973), such a differential rotation in a
strongly ionized medium is sufficient to maintain a magnetic dynamo,
even in the absence of convection (see also Mestel 1999; Goedbloed
\& Poedts 2004). This would generalize the notion of a stellar
dynamo as discussed by Parker (1975) for the convective envelope.
This specific approach is however critiqued in recent papers by
Denissenkov \& Pinsonneault (2007) and Zahn et al. (2007).

In the rest of this note, we do not adopt a specific dynamo model.
However, we do assume that toroidal magnetic fields may exist in
giant stars and that buoyant flux tubes will be generated in the
radiative layers below the envelope. In fact, whatever the correct
model for an effective stellar dynamo might be, it is sufficient for
us to know that such a dynamo operates rather ubiquitously in low
mass stars (see e.g. Baliunas and Vaughan 1985) and to derive from
the solar scenario the suggestion that it is generated in radiative
layers (see Fan 2004, and discussion below).

Detailed dynamical models for the Sun by Zahn (1992) early suggested
a change of about 30\% in the equatorial rotational speed from the
inner core to the surface. These models were important
anticipations. However, they did not account for a magnetic dynamo
and are today no longer quantitatively supported by helioseismology
results. These last, in particular after the SOHO measurements, have
established that the convective envelope of the Sun has at its base
a region in the nominally radiative zone, roughly 0.04 $R_{\odot}$
and $\sim$ 0.01 \msb thick (called the {\it tachocline}) in which
rotation gradually passes from rigid (in the central zones) to
strongly differential (in the convective outer layers). For the Sun
the angular velocity of the core is intermediate between the
equatorial and high-latitude surface spin rates. For recent reviews
on these subjects see Fan (2004), Miesch (2005).  Models of
radiative dynamos, despite the specific problems they may present,
have the merit of accounting for helioseismology results
(Eggenberger et al. 2005) and of explaining how the rigid rotation
of the core can be maintained, despite the contrary arguments by von
Zeipel (1924), if the dynamo soaks up the transported angular
momentum as quickly as meridional circulation can provide it.

Observations of the Sun's magnetically-active regions, especially at
intermediate and low latitudes, are now understood in the framework of
an accepted paradigm. The current interpretation (Fan 2004, Miesch
2005) includes the following elements: i) a dynamo mechanism is
generated in the tachocline (a nominally radiative region),
maintaining toroidal fields, probably organized as isolated, thin flux
tubes (Spruit 1981); ii) The tube profiles are greatly modified into
various undulatory shapes (Alfv\'en waves), sometimes evolving into
instabilities (Spruit \& van Ballegooijen 1982). Among them, kink-type
modes evolve into $\Omega$-shaped loops with strong radial components
(Parker 1974; 1994). iii) Such buoyant loops penetrate the convective
layer and then emerge forming the corona; photospheric active regions
and sunspots are cross sections of bundles of many flux tubes
(``spaghetti model'') and are the roots of coronal loops. Magnetic
buoyancy is compensated by complex downflows, providing a sort of
asymmetric circulation (Spruit 1997).  We will assume that a layer
analogous to the tachocline exists in RGB and AGB stars, and that flux
tubes will rise through this region as they do in the sun.

In the Sun, magnetic field intensities in flux tubes are of a few
$\times$ 10$^3$ G in sunspots, which emerge from the photosphere.
These fields must considerably exceed 10$^5$ G deep in the radiative
tachocline from which they come (Rempel et al. 2000; Sch\"ussler and
Rempel 2002). Certainly, the fields exceed 10$^4$ G (Fan 2004) in
active regions at the bottom of the convective layer. For the Sun,
the value of a few $\times$ 10$^4$ G corresponds to equipartition of
energy between the magnetic ($B^2/ 8\pi$) and convective-kinetic
(1/2 $\rho v_c^2$) forms in the innermost convective zones. For the
typical density of 0.15 g/cm$^3$ (Bahcall et al. 2006), this implies
that, at equipartition, the convective velocity $v_c$ averages at
tens of meters per second.

The observed values of the solar magnetic flux, which are at the
level of a few $\times 10^{21}$ Maxwell in ``small'' active regions
(bundles of many filaments, see e.g. Zwaan 1987, Table 2), together
with local field intensities of a few $\times 10^4$G, imply that
magnetized zones of $\sim 1000 - 2000$ km in radius exist deep in
the convective layer. We shall use this estimate of flux tube
dimensions in calculations below.

\subsection{Investigating the links between Magnetic Buoyancy and CBP}

Subadiabatic zones below the convective envelope, qualitatively
similar to the solar tachocline (i.e. regions where the heat transport
occurs through radiative processes, even in the absence of a strong
chemical stratification) can be found in low mass stars after they
have reached the red giant branch, both in the final thermally-pulsing
phases of the AGB (cf. Nollett et al. 2003) and in preceding RGB
stages, which are of much longer duration. The necessary condition is
that the advancing H shell has erased the chemical discontinuities
left behind by core H-burning and by the first dredge-up. In all such
zones, where the molecular weight gradient is close to zero, stability
against matter circulation is not guaranteed (Fricke \& Kippenhahn
1972; see also Collins 1989, Chapter 7) and cool bottom processes
might occur.  Results using the code of Nollett et al. (2003) show
that the data can be described by a mass mixing rate of roughly
$10^{-6}$\ms/yr (AGB case) or $4\times 10^{-8}M_\odot$/yr (RGB case)
through the top 80--90\% by mass (97--99\% by radius) of the region
between the hydrogen-burning shell and the base of the convective
envelope.  (These numbers will be developed in more detail in Section
3.)

Here we seek to ascertain whether magnetic buoyancy is a plausible
option to transport matter in red giants, and to provide the mass
circulation rate $\dot M$ required by CBP.  The requirements from
CBP are only a rate of transfer and the maximum temperature of
burning. One of the rate-limiting conditions for mass transfer
occurs at the boundary between the convective envelope and the
radiative zone.  The fraction of mass ($f_m$) at the upper part of
the radiative zone made up of flux tubes must be small so that the
stellar structure is not greatly disturbed.  The velocity of
transport across the boundary must also be compatible with available
estimates.  We may write the mass flow rate as:
\begin{equation}
\dot M \simeq 4\pi r^2 \rho c_s f_m f_t,  %%\eqno(1)
\end{equation}
where $c_s$ is the velocity of sound; $r$ is the radius at the
convective envelope base; $\rho$ is the mass density; $f_t$ is the
ratio between the velocity of transport across the boundary to that of
sound.  Using the local values of the density and of the sound speed
obtained from the stellar model, we will first verify that the
$\dot{M}$ required for CBP corresponds to small $f_m$ and $f_t$, so
that the fraction of mass at the upper radiative zone interface which
is provided by buoyant flux tubes is $f_m \ll$ 1 and that the
transport velocity across the boundary ($f_t \times c_s$) is
reasonable.  We will further check whether the magnetic field
requirements of our flux-tube model is consistent with small $f_m$ and
$f_t$.

We may also derive from $\dot{M}$ the rate $\dot{N}_t$ at which rising
flux tubes must arrive at the convective envelope.  At the initial
position $r_0$ above the H shell let a flux tube be a torus with an
initial radius $a_0$, length $l_0 = 2 \pi r_0$, and a magnetic field
intensity $B_0$. The volume of the tube is 2$\pi^2 r_0 a_0^2$
cm$^3$. At the local density $\rho_0$ this corresponds to a mass of $2
\pi^2 r_0 \rho_0 a_0^2$. During an evolutionary phase of duration
$\Delta t_{ph}$, enough tubes must be deposited into the envelope to
guarantee the isotopic changes provided by CBP. In order to achieve
this, the convective envelope itself must be mixed, within the
available time $\Delta t_{ph}$, with an amount of H-burning-processed
material equal to a fraction $\alpha$ of its mass $M_E$ (in solar
units).  The number $N_t$ of the tubes must be therefore of the order
of $\alpha M_E \times (1.989 \times10^{33}\ \mathrm{g})/(2 \pi^2 r_0 \rho_0
a_0^2)$. The rate at which these flux tubes must reach the boundary
layer is thus:
\begin{equation}
\dot N_t = {{\alpha M_E (1.989 \times 10^{33}\ \mathrm{g})} \over {2 \pi^2 r_0
\rho_0 a_0^2 \Delta t_{ph}}}
= {{\dot{M}} \over {2 \pi^2 r_0
\rho_0 a_0^2}} = {{\hat v}\over {\Delta r}}%%\eqno(2)
\end{equation}
where $\hat v$ is the average velocity of the tubes.  $\Delta r$ is
the distance from the processing zone to the convective envelope.

We can now explore the implications of the mass circulation
requirements on the magnetic fields near the H-burning shell. We
assume, after Parker (1974), that there exists a fractional shift $s$
in density due to the magnetic pressure from the matter density $\rho$
outside a flux tube to the matter density $\rho^\prime$ within the
flux tube. This shift is given by:
\begin{equation}
{(\rho - \rho') \over \rho} = {(B^2/8 \pi) \over P} \equiv s
%%\eqno(3)
\end{equation}
Applying equation (3) throughout the motion of a flux tube assumes
that the interior of the tube is always at the local temperature.
This condition is plausible only if the buoyancy velocity is
sufficiently small (below few km/sec, cf Parker 1974).

Balancing the buoyancy force per unit length on the flux tube ($F =
\pi a^2 g(r) (\rho -\rho')$) with the drag force per unit length
($1/2 C_D \rho a v^2$), we obtain $v^2 = (2 \pi/C_D) \times g(r)
a(r) \Delta \rho/\rho$, where $g(r)$ is the acceleration of gravity
at the position $r$ and $C_D$ is the aerodynamic drag coefficient.
Note that there is little mass in the radiative layer of either the
RGB or AGB star, so that $g(r)/g_0 \sim (r_0/r)^2$. Then:
\begin{equation}
v^2(r) = {{2 \pi a(r) g_0}\over C_D} \left({r_0 \over r}\right)^2
{{\Delta \rho}\over \rho} %%\eqno(4)
\end{equation}
Assuming conservation of magnetic flux $ B(r) = {B_0 a_0^2/a(r)^2}$
and conservation of mass $ a(r)^2 =
{\rho_0 a_0^2 r_0 /[\rho(r) r]}$  within the flux tube,
one derives:
\begin{equation}
B(r) = B_0 {{\rho(r) r} \over{ \rho_0 r_0}} %%\eqno(5)
\end{equation}
Hence equation (3) becomes:
\begin{equation}
{\Delta \rho \over \rho} = \left({B_0^2 \over 8 \pi}\right) \left({r
\over r_0}\right)^2 \left({\rho \over \rho_0}\right)^2 \times {1
\over P} %%\eqno (6)
\end{equation}
The velocity can therefore be expressed as:
\begin{mathletters}
\begin{equation}
v(r) = {1 \over 2} \left({\rho(r) \over \rho_0}\right)^{3/4}
\left({r \over r_0}\right)^{-1/4} \left({g_0 a_0 \over
C_D}\right)^{1/2} \left({B_0 \over \sqrt{P(r)}}\right) %%\eqno(7a)
\end{equation}
or:
\begin{equation}
{v(r) \over v_0} = \left({r \over r_0}\right)^{-1/4}\left({\rho(r)\over \rho_0}\right)^{3/4}
\left({P(r)\over P_0}\right)^{-1/2} %%\eqno (7b)
\end{equation}
\end{mathletters}
Concerning the effective (average) velocity of the buoyancy motion
over the trajectory, $\hat{v}$, let the time of transport of an
individual flux tube across the radiative layer be $t_T$; then
$\hat{v} t_T = \Delta r$ ($\Delta r$ being the distance from the
starting position near the H shell, $r_0$, to the base of the
convective envelope, $r_E$).  From equation (7a) one can then
compute:
$$
t_T = \int_{r_0}^{r_{E}} {dr \over v(r)}
$$
And finally:
\begin{equation}
\hat{v} = {{\Delta r} \over {\int_{r_0}^{r_{E}} {dr \over v(r)}}}
%%\eqno(8)
\end{equation}

The above rough approximations are sufficient for order-of-magnitude
estimates applied to specific physical conditions inside evolved red
giant stars.  In the next Section, we will estimate $f_m$, $f_t$, and
$\hat{v}$ from requirements on $\dot{M}$ and $T_P$.  From $\hat{v}$,
we will use equations (7) and (8) to infer the magnitude of the
magnetic field required at the bottom of the CBP circulation in order
for buoyant flux tubes to carry it.

\section{Requirements on Magnetic Buoyancy for Evolved Red Giants}

In order to make some quantitative estimates of the magnetic fields
that would account for CBP nucleosynthesis, we shall consider, as a
reference, the case of a 1.5 \msb red giant, with a metallicity half
the solar one and an internal structure as computed by Straniero et
al (1997) and Busso et al. (2003), including mass loss with the
parameterization by Reimers (1975). In this model we shall examine
first the AGB situation. Here the occurrence of CBP has quite
stringent requirements, as it must affect nuclei up to Mg-Al in a
rather short interval of time and must reduce the carbon isotopic
ratio $^{12}$C/$^{13}$C in the envelope in competition with the
ongoing, recurrent enhancement of the $^{12}$C abundance provided by
the third dredge-up. According to Nollett et al. (2003) the whole
range of the observations can be accounted for if circulation rates
are in the interval 10$^{-7}$ to 10$^{-5}$ \ms/yr. We shall choose
$\dot M = 10^{-6}$ \ms/yr as a representative case.

A typical structure of the radiative zone above the H-burning shell,
in the AGB phase of our model star of initially 1.5 \ms, is
illustrated in Table 1(a). The values given are typical for the
interpulse periods (those shown are for the period between the 8th and
the 9th occurrence of the third dredge-up). The same physical
properties are also plotted in Figure 1. The envelope mass $M_E$ and
the radius $R_0$ of the model star at the phase considered (where mass
loss has reduced the total mass to 1.207 \ms) are shown in the top of
Table 1(a).  The total time available for the thermally-pulsing AGB
stage, summing all interpulse periods (and excluding the relatively
short duration of dredge-up episodes) is $\Delta t_{ph} \simeq 2\times
10^6$ yr.  The depth (temperature) to which the circulating mass must
reach is that characterized by $\log ~T_P = \log ~T_H -0.1$ (Nollett
et al. 2003).  Here $T_H$ is the temperature of the hydrogen-burning
shell and $T_P$ is the temperature at the maximum depth at which
matter is transported.

The conditions pertinent to the RGB stages where we consider CBP
correspond to the initial state where the advancing H-burning shell
has erased the chemical discontinuities left behind by core-H burning
and by the first dredge-up. On the RGB there is no $^{12}$C added to
the envelope, as occurs instead with the repeated dredge-ups on the AGB.
An RGB structure used for that stage is shown in Table 1(b), in the
same format as before. The structure is also sketched in Figure 2. The
depth to which the circulating mass must reach is taken at $\log~T_P = 7.4$,
sufficient for a large fractional change in the $^{13}$C content.
The total time on the RGB for providing the nuclear processed material
is $\Delta t_{ph}$ = 4.3$\times$10$^7$ yr.

For the RGB stages, CNO isotopes (and Li) are known to be affected
by CBP, the observed $^{12}$C/$^{13}$C ratio going down from an
initial value of $\sim$ 25 to about 10-13 (in population I stars).
The RGB evolution thus provides $^{13}$C enrichment and some
$^{12}$C destruction ($\sim$ 30\%). Contrary to the AGB case, no
concurrent $^{12}$C enrichment by dredge-up is present. Moreover,
the time available for mixing is much longer than for the AGB so
that we expect much less challenging conditions for the required
magnetic fields.

As a comparison, the physical properties for the Sun below the
convective envelope are shown in panel (c) of Table 1. Note that in
the Sun the tachocline mass is $\Delta M =$ 7.8$\times$10$^{-3}$ \msb
and its thickness is $\Delta R$ = 0.04 $R_{\odot}$. The mass of the
overlying convective envelope is 0.02 \ms. In contrast, for the AGB
and RGB cases the radiative layer has a mass from a few
$\times$10$^{-4}$ to a few $\times$10$^{-3}$ \ms, and a much larger
thickness of $\Delta R \sim$ 0.79 $-$ 0.86 $R_{\odot}$.  In addition, the mass
of the convective envelope is obviously very different from the Sun in
these extended red giants, and ranges from 0.5 to more than 1 \ms.  Thus,
the distances traveled by the hypothesized flux tubes in both the RGB
and AGB stars is $\sim 20$ times larger than for the sun, and the overlying
envelopes are also much larger.
%\clearpage
\begin{figure}[t]
\centerline{{\includegraphics[width=9cm]{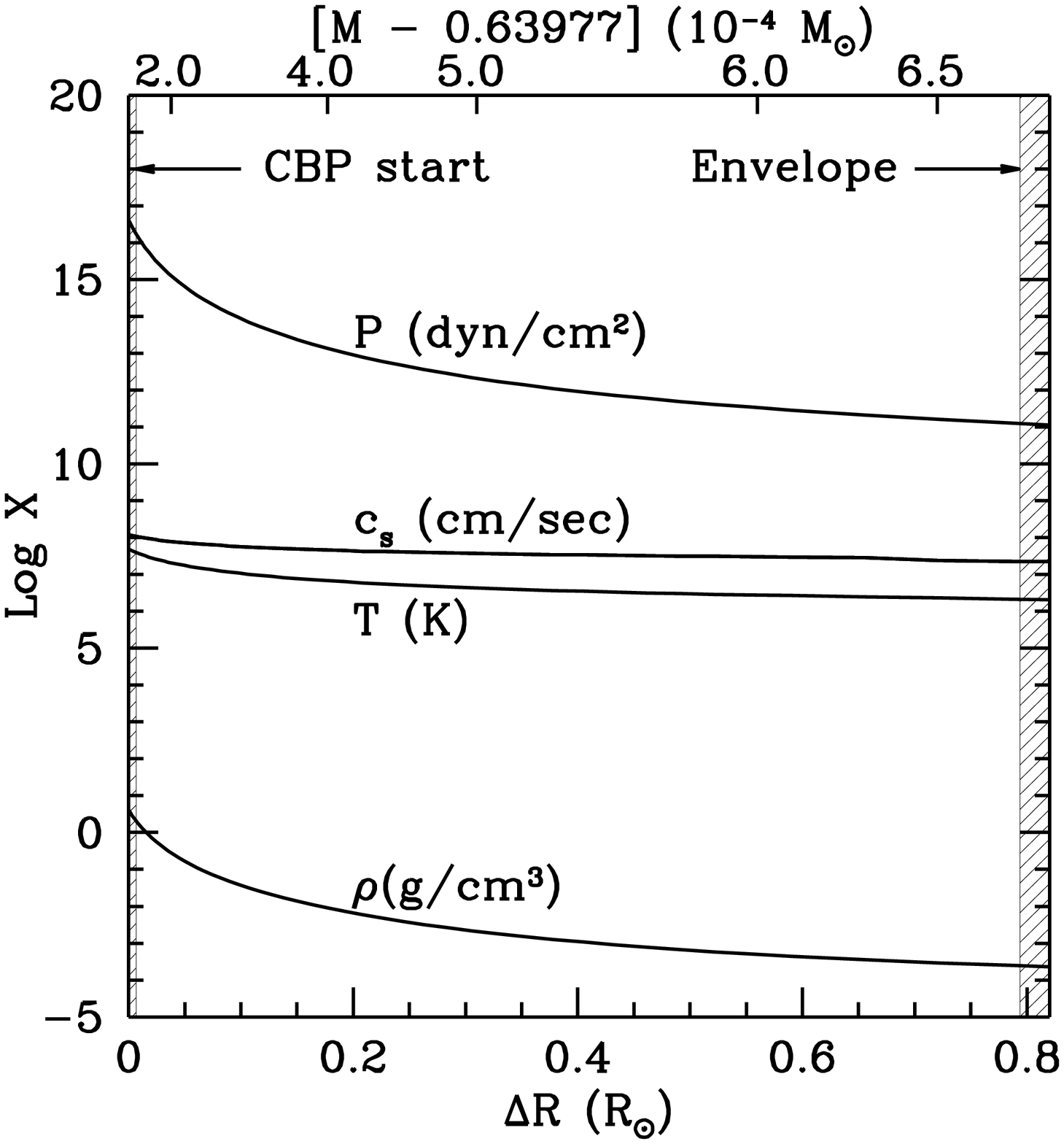}}}
    \caption{The behavior of pressure, density, temperature and sound speed as
    a function of the distance (expressed both in radius and in mass) from the
    point where we assume CBP starts. The plot shows the radiative layers
    below the convective envelope, for the thermally pulsing AGB phases of a
    1.5 \msb star with half solar metallicity (see text).}
              \label{Fig1}%
    \end{figure}

\begin{figure}[t]
\centerline{{\includegraphics[width=9cm]{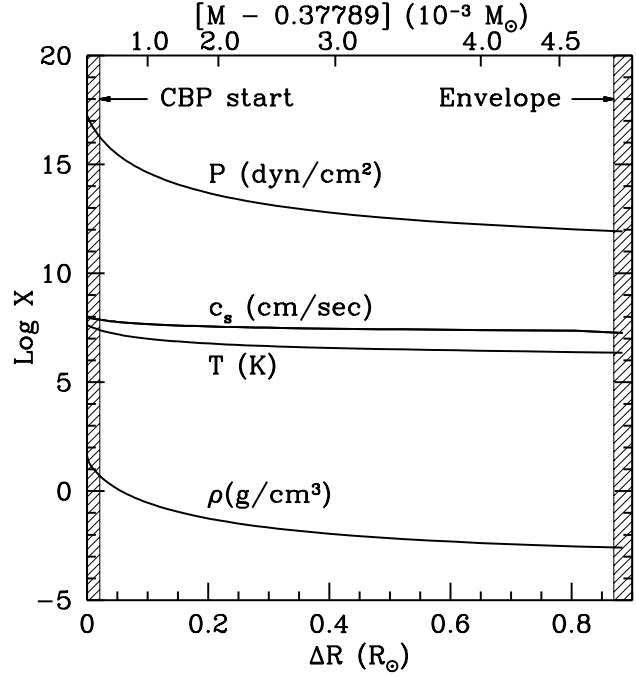}}}
    \caption{A typical structure of the radiative layers below the convective
    envelope on the RGB, characterizing the stages after the luminosity bump and up to
    core-He ignition (see text).}
              \label{Fig2}%
    \end{figure}
%\clearpage

\begin{deluxetable}{ccccccc}
\tablecolumns{7}
\tablewidth{0pt}
\tablecaption{Relevant parameters in the sub-convective layers}
\tablehead{
\colhead{Stellar ~Zone} &
\colhead{$M/M_0$} &
\colhead{$r/R_0$} &
\colhead{$P (dyn/cm^2)$} &
\colhead{$T (K)$} &
\colhead{$\rho (g/cm^3)$} &
\colhead{$c_s (cm/sec)$}
}
\startdata
\cutinhead{\bf (a) AGB phase\tablenotemark{1}. $M_0 =$ 1.207
$M_{\odot}$, $M_E =$ 0.567~ $M_{\odot}$, $R_0 =$ 331 ~$R_{\odot}$,
$\Delta t_{ph} = 2\times 10^6$ yr)}

\noindent H shell position & 0.529999 & 6.59 10$^{-5}$  &  1.42 10$^{17}$  & 6.34 10$^{7}$ & 17.71 & 1.07 10$^{8}$\\
  \hline
\noindent  Max. CBP penetration\tablenotemark{2} & 0.530049 & 8.57 10$^{-5}$  &  4.24 10$^{16}$  & 4.92 10$^{7}$ & 4.13 & 1.19 10$^{8}$\\
  \hline
\noindent  Top Radiative Zone & 0.530519 & 2.26 10$^{-3}$  &  1.60 10$^{11}$  & 2.32 10$^{6}$ & 2.92 10$^{-4}$ & 2.75 10$^{7}$\\
  \hline
\noindent  Bottom Conv. Zone & 0.530558 & 2.46 10$^{-3}$  &  1.29 10$^{11}$  & 2.17 10$^{6}$ & 2.48 10$^{-4}$ & 2.67 10$^{7}$\\

\cutinhead{\bf (b) RGB phase\tablenotemark{1}. $M_0 =$ 1.499 $M_{\odot}$, $M_E =$
1.115~ $M_{\odot}$, $R_0 =$ 53 ~$R_{\odot}$, $\Delta t_{ph} = 4.3
\times 10^7$ yr.)}

\noindent H shell position & 0.251564 & 5.41 10$^{-4}$  &  1.87 10$^{17}$  & 4.19 10$^{7}$ & 43.05 & 2.52 10$^{8}$\\
  \hline
\noindent  Max. CBP penetration\tablenotemark{3} & 0.252093 & 9.34 10$^{-4}$  &  1.87 10$^{16}$  & 2.51 10$^{7}$ & 5.19 & 7.44 10$^{7}$\\
  \hline
\noindent  Top Radiative Zone & 0.255296 & 1.58 10$^{-2}$  &  1.05 10$^{12}$  & 2.42 10$^{6}$ & 2.93 10$^{-3}$ & 2.32 10$^{7}$\\
  \hline
\noindent  Bottom Conv. Zone & 0.255528 & 1.72 10$^{-2}$  &  8.49 10$^{11}$  & 2.26 10$^{6}$ & 2.55 10$^{-3}$ & 2.23 10$^{7}$\\

\cutinhead{\bf (c) The present Sun\tablenotemark{4}, $M =
M_{\odot}$, $R = R_{\odot}$}

\noindent  ~~Tachocline base~~~~~~~~~ & 0.972383 & 0.6880  &  7.09 10$^{13}$  & 2.36 10$^{6}$ & 0.219 & 2.45 10$^{7}$\\
  \hline
\noindent  Tachocline top & 0.980219 & 0.7279  &  4.34 10$^{13}$  & 2.01 10$^{6}$ & 0.155 & 2.18 10$^{7}$\\
  \hline
\noindent  ~Bottom Conv. Zone~ & 0.980280 & ~~0.7282~~  &  4.32 10$^{13}$  & ~2.00 10$^{6}$ & ~0.150 & 2.15 10$^{7}$\\

\enddata
\tablenotetext{1}{From a stellar model with initial mass $1.5M_\odot$ and initial
metallicity one-half solar.}
\tablenotetext{2}{Defined as the layer where $\log~T = \log~T_P = \log~T_H - 0.1$ (Nollett et al 2003).}
\tablenotetext{3}{Assumed as the layer where $\log~ T_P$ = 7.4 (Substantial production of $^{13}$C).}
\tablenotetext{4}{From Bahcall et al. (2006), (data published electronically)}

\end{deluxetable}

%\clearpage

\subsection{The AGB case}

Adopting, at the radiative-convective boundary of the AGB stages,
$\dot M =$ 10$^{-6}$ \ms/yr in equation (1), together with the local
density of 2.48$\times 10^{-4}$ g/cm$^{-3}$, and with the local
velocity of sound of 2.67$\times$10$^7$ cm/sec (Table 1a), from
equation (1) one gets, for material transport across the convective
envelope border, $f_m f_t \simeq $ 2.33$\times$ 10$^{-7}$.

We do not have an a priori estimate for the buoyancy velocity. Fan
(2004) suggests that, near the convective border of the Sun, it can
be $v \simeq 10^{-3} |\delta|^{-1}$ cm sec$^{-1}$, where $\delta$ is
the difference between the logarithmic thermal gradient ($d ~\log~ T
/ d ~\log ~P$) and the adiabatic one. As this difference becomes
typically $\delta \sim - 10^{-5}$ in the last subadiabatic layers
below the convective solar envelope, the value of $v$ is close to 1
m/sec. Values of this order would imply $f_t \cong 3.6 \times
10^{-6}$, and $f_m \cong 0.07$. As we shall see later (Figure 3),
from our formulae the AGB buoyancy velocity at base of the
convective envelope turns out to be higher, typically 1 km/sec. In
this case $f_t \cong 3.6 \times 10^{-3}$, and $f_m \cong 7 \times
10^{-5}$. Thus, in any case only a small fraction of the mass in the
top of the radiative zone must be from flux tubes.

According to Nollett et al. (2003), CBP must connect the envelope with
internal zones where the maximum temperature $T_P$ is as high as $\log
~T_P - \log ~T_H = - 0.1$. Transport from this level into the
convective envelope provides an adequate mass of processed material if
$\dot M \sim 10^{-6}$ \ms/yr and $\alpha \simeq 1$.

For the AGB case shown in Table 1(a), and assuming the flux tubes to
be toroidal at all $r$, the ratio $r_0/r$ between the innermost
region where CBP must penetrate and the base of the envelope is
$\sim 1/28.7$. The density ratio is $\rho_0 /\rho \sim 1.66~10^4$.
This implies that $a_0/a \sim 1/24$. If we assume that the flux
tubes have the same size of $\sim (1000 - 2000)$~km at the base of
the convective envelope as in the Sun (as discussed in section 2.1),
we obtain an initial tube radius of $a_0 \sim $ (42 - 83)~km in the
zone where CBP starts. We adopt $a_0 = 65$ km for illustration. For
an envelope mass of $M_E = 0.567 $ \ms, and with the parameters of
Table 1(a), equation (2) yields $\dot N_t \simeq 2.6 \times 10^{-6}
sec^{-1}$ ($\sim$ 7 per month). The corresponding average velocity
to deliver a parcel of matter to the envelope (at $r = 5.7\times
10^{10}$ cm) is $\hat v \sim$ 1.45 km/sec if only one flux tube
rises at a time. This relatively low velocity justifies {\it a
posteriori} our assumption of thermal equilibrium between the
magnetized flux tube and the environment. If any form of magnetic
diffusion or phase mixing (Spruit 1999) were to occur, then a
fraction of the flux tubes originally formed would not reach the
convective zone interface, so the net rate of generation of flux
tubes would have to be greater. If many ($n$) flux tubes form in the
processing region at essentially the same time, then the velocity is
decreased by a factor $n$. The number of flux tubes that can be
generated simultaneously just above the H shell is not known.
Insofar as this number is not much greater than unity, the
conclusions are not strongly dependent on $n$. In all the following
discussion we shall in this paper take $n$ =1. The common modelling
scheme called the {\it thin flux tube approximation} (Sch\"ussler
1977, Spruit 1981), which we have followed, essentially assumes that
flux tubes remain as individual entities deep in the stellar
interiors. We note in contrast that the magnetized zones in the
solar envelope to which we make reference in deriving $a_0$ are
bundles of many individual filaments. We cannot ascertain now
whether these bundles are formed as groups of separate filaments or
as individual flux tubes, but we note that turbulence in the
envelope is in itself a very likely and efficient cause for locally
shredding magnetic tubes into many perturbed filaments (cf Stothers
2004).

If the average velocity necessary to provide processed matter to the
envelope at the required rate is of the order estimated above, then
we can derive the value $v_0$ of the maximum velocity (the one
pertaining to the innermost layers, near the H shell). For the AGB,
using equations (8) and (7a), and computing numerically the
integrand gives $v_0 \sim$ 6.8 km/sec, for $\hat v = 1.45$ km/sec:
the function of equation 7(b) is shown in Figure 3 (continuous
line). Using the pressure of 4.24$\times 10^{16}$ dyne/cm$^2$ in the
CBP starting zone, we obtain a magnetic field value in the innermost
layers of $B_0 \simeq ~ 2.3\times10^7 C_D^{1/2}$.

The value of $C_D$ is uncertain. At high Reynolds numbers it is
usually assumed to be close to unity, after Batchelor (1967), but the
range of the possible values might extend down to about 0.04 (Hans
Hornung, private communication), if one considers that the boundary
between the flux tubes and the surrounding medium permits slip. The
lowest necessary fields would therefore correspond to $C_D^{1/2} =
0.20$. We can note that the stored fields in the solar radiative
layers do not largely exceed the equipartition value (Moreno Insertis
1986). If equipartition is applicable for AGB stars, the condition
$B_0^2/8 \pi$ = 1/2 $\rho_0$ $v_0^2$ and the velocity $v_0$ give $B_0
= 5\times 10^6$ G. This formally corresponds to a value of $C_D =
0.05$. As $C_D$ appears under a square root, changing its value has a
moderate effect on the field (for example doubling the above choice,
i.e. adopting $C_D = 0.1$, would imply a field $B_0$ = 7$\times$10$^6$
G, still rather close to equipartition).

With due consideration for the uncertainties in the above
calculations, it is therefore clear that strong fields are needed
near the H shell to drive CBP, at the level of several Megagauss
(MG) in the innermost zones reached by mass circulation. At the base
of the convective envelope, using equation (5) and $B_0 = 5\times
10^6$ G we get $B_{E} \simeq$ 9$\times$ 10$^3$ G. This is similar to
the fields found in the deep layers of the solar convective
envelope. 

%\clearpage
\begin{figure}[t]
\centerline{{\includegraphics[width=9cm]{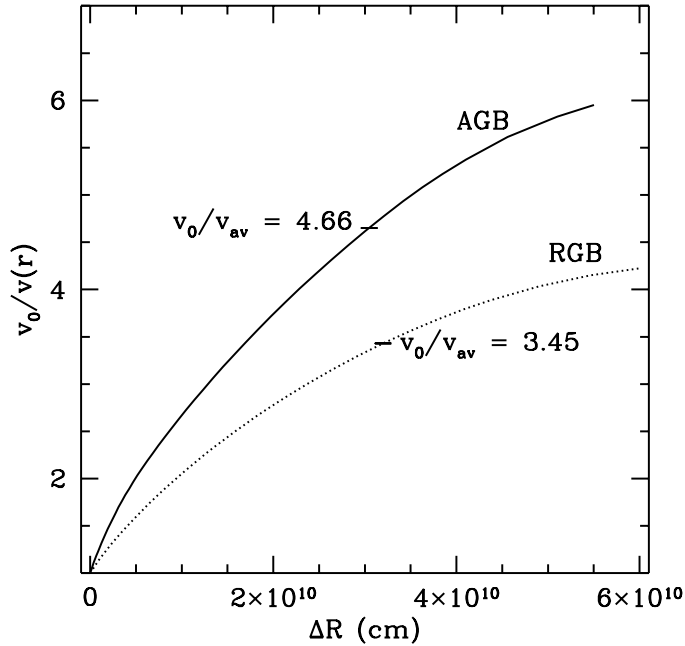}}}
    \caption{The buoyancy velocity $-$ expressed, as in equation (7b), through
    the ratio between the maximum value and the local value at any $r$ $-$ as a function
    of the distance from the point where we assume CBP starts. The ratio of the
    initial velocity and the average velocity is shown for the two cases. See text.}
              \label{Fig3}%
    \end{figure}

\begin{figure}[t!]
\centerline{{\includegraphics[width=9cm]{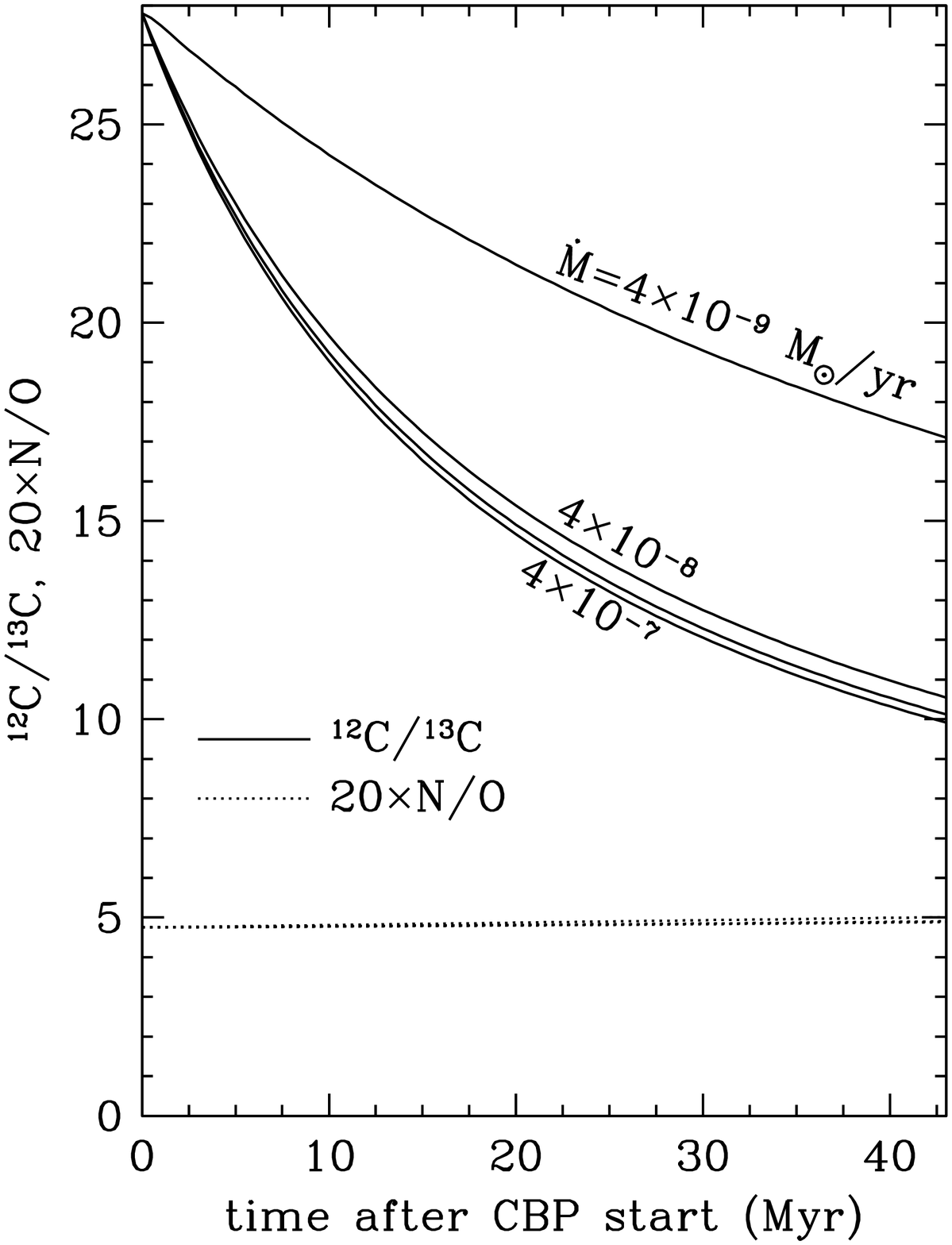}}} \caption{CBP
model results for the RGB stellar structure of Fig. 2, four
different mixing rates $\dot{M}$, and an assumed maximum mixing
depth with $\log (T_P/T_H)=-0.22$.  Curves show $^{12}$C/$^{13}$C
and N/O as functions of time after start of CBP.  Three values of
$\dot{M}$ are shown next to the corresponding $^{12}$C/$^{13}$C
curves; the fourth is $10^{-7}M_\odot$/yr.  For this mixing depth,
$^{12}$C/$^{13}$C constraints are satisfied as long as $\dot{M}
\gtrsim 10^{-8}M_\odot$/yr.  All four curves for N/O lie on top of
each other.  Higher $T_P$ would require $\dot{M}$ near $4\times
10^{-8}M_\odot$/yr to satisfy the $^{12}$C/$^{13}$C constraints and
would show N production.}
\label{Fig4}%
\end{figure}
%\clearpage
\subsection{The RGB Case}

Unambiguous evidence for CBP in first-ascent red giants is offered
by the ratio $^{12}$C/$^{13}$C, which decreases more sharply than
can be accounted for by first dredge-up. Average values of $\dot M$
and $T_P$ suitable for the RGB can then be estimated from inspection
of the stellar model and from simple considerations. We refer to
late RGB stages, after the H-shell has erased previously established
chemical discontinuities. From this moment on, the layers below the
convective envelope have a rather homogeneous composition, so that
the chemical stratification does not seriously hamper matter
circulation. In the model of Table 1(b), the duration of this phase
is about 4.3$\times$10$^7$ yr. The convective envelope mass
decreases over this time period from 1.203 $M_{\odot}$ to 1.001
$M_{\odot}$, with an average value of $M_E$ = 1.11 \ms. CBP must
bring the $^{12}$C/$^{13}$C ratio of this convective envelope down
to $\sim 10 - 13$, starting from the typical value of 25 left behind
by first dredge-up. This must be obtained by mixing the envelope
with a total mass $M_c$ of circulated material greatly enriched in
$^{13}$C (possibly approaching the CN cycle equilibrium of
$^{12}$C/$^{13}$C$\lesssim  4$). This estimate implies $\alpha \sim
1$ and therefore $\dot M$ values larger than 10$^{- 8}$ \ms/yr, over
the time interval available. A more detailed calculation of the
evolution of the envelope for this RGB star is shown in Figure 4 for
$\log~T_P$ = 7.4, corresponding to $\log (T_P/T_H)=-0.22$, and $\dot
M$ in the range $4\times 10^{-9}$ to $4\times 10^{-7}$ \ms/yr. These
calculations were carried out using the CBP code of Nollett et al.
(2003). It can be seen, from the final carbon isotopic ratio
obtained, that the estimated value of $\dot M$ is correct. As an
average, for the RGB case we adopt $\dot M = 4\times 10^{-8}$ \ms/yr
and $\log~T_P$ = 7.4. It is to be noted that for this choice the N/O
ratio is unaffected. ($^7$Li, not shown in the figure, is instead
extensively destroyed).

First we note that the condition established in equation (1), that
$f_t$ and $f_m$ be small, is well fulfilled. Their product is $f_t
\times f_m$ = 8.7$\times$10$^{-10}$. For a transport velocity of
$\sim$ 100 cm/sec near the envelope base (Fan 2004 and Sec. 3.1
above), this gives $f_m \cong 2\times 10^{-4}$. As we shall see, by
estimating the buoyancy velocity at the envelope base through equation
(7b) one gets a higher value of $\sim$ 10 m/sec (Figure 3). In this
case $f_m \cong 2\times 10^{-5}$. Again, for any reasonable choice of
the transport velocity, only a minimal fraction of the mass needs to
be from flux tubes at the radiative-convective boundary.

We may now directly obtain the results for the RGB as compared to the AGB. The
condition for the average velocity (equation 2) applies both to
the AGB and to the RGB case. Hence, indicating with subscripts (A) and (R) the AGB and
RGB cases, respectively:
\begin{equation}
{\hat v_{R} \over \hat v_{A}} = {{{\Delta t_{ph,A} \Delta r_{R} (\alpha M_E)_{R}}
r_{0,A} \rho_{0,A}} \over {\Delta t_{ph,R} \Delta r_{A}(\alpha M_E)_{A} r_{0,R}
\rho_{0,R}}} \left({a_{0,A}}\over{a_{0,R}}\right)^{2} %%\eqno(9)
\end{equation}
with $\alpha_{R} =1.48$ from $\dot{M}=4\times 10^{-8}M_\odot$/yr as found in
the CBP calculation for the RGB. This yields:
\begin{equation}
{\hat v_{R}\over \hat v_{A}} = 6.78\times10^{-2} \left(
{a_{0,A}\over a_{0,R}}\right)^2 %%\eqno(10)
\end{equation}
A similar relation (derived from equation 7a) holds for the
maximum velocity [$v_0$ = 1/2 ($g_0 a_0/C_D)^{1/2}(B_0/P_0^{1/2})$], so that if $C_D$ is
equal in both cases:
\begin{equation}
{v_{0,R}\over v_{0,A}} = {{(g_0 a_0)_{R}^{1/2}B_{0,R} P_{0,A}^{1/2}}\over
{(g_0 a_0)_{A}^{1/2} B_{0,A} P_{0,R}^{1/2}}} %%\eqno(11)
\end{equation}
Taking the ratio of equations (10) and (11), together with the relationship between
$\hat v$ and $v_0$ shown in Figure 3, and using the stellar code parameters of Table
1 (a, b), we obtain the following scaling law:
\begin{equation}
{B_{0,R}\over B_{0,A}} =  0.075 \left(a_{0,A} \over
a_{0,R}\right)^{5/2} %%\eqno(12)
\end{equation}
\noindent It is evident that, for the same initial flux tube
dimensions, $B_{0,R}/B_{0,A} \lesssim 1/10$. The $5\times 10^6$ G
field for the AGB stages would in particular correspond to a field of
$\sim 3.8\times 10^5$ G for the RGB stages.  From Equations (10) and
(2), this corresponds to $\hat{v}=98$ m/sec and $\dot{N}_t =
1.63\times 10^{-7}$ sec$^{-1}$ ($\sim 5$ times per year) during the
RGB phase.

If we assume instead that $a$ at the bottom of the convective envelope
is the same (1000 -- 2000 km) on the RGB as we assumed for the AGB,
then the parameters of Table 1(b) give $a_0 \sim 100$ to 200 km, for
an average of $\sim$ 150 km.  Since $\dot{M}_{RGB} = 4\times
10^{-8}M_\odot$/yr, the rate of flux tube generation for this case,
expressed by equation (2), is $\dot{N}_t = 3 \times 10^{-8}$
sec$^{-1}$ ($\sim$ once per year) and the velocities are $\hat{v}\sim
18$ m/sec and $v_0 \sim 62.4$ m/sec. Using these values of the
velocity and $C_D \sim 0.05$ to infer an estimate for $B$ in the
deepest layers affected by CBP (equation 7a), we get $B_0 \sim
4.8\times 10^4$ G, close to the equipartition value ($4.3\times 10^4$
G). [Note that the same values for the velocities and the magnetic
field $B_0$ can also be derived by equations (10), (11) and (12),
adopting $a_0 = 150$ Km]. The estimate of $B_0 = 4.8\times 10^4$ G
gives $B \sim 440$ G at the bottom of the convective envelope from
flux conservation.

The above results for the AGB and RGB fields and buoyancy velocities
are summarized in Table 2.  Velocities in all cases are small enough
that thermal equilibrium with the environment should be achieved.
There is a significant difference in the pressure scale height
(hence in any scale height for mixing) at the CBP position for the
RGB as compared to the AGB. Hence arguments by Vishniac (1995), in
particular his equation (61), would indicate that the larger value
of $a_0$ in the case RGB-2 of Table 2 may be the correct one for the
RGB.  For either value of $a_0$, the result is that substantial
magnetic fields are also required by CBP on the RGB, though at far
lower levels than for the AGB case.

%\clearpage
\begin{table}[t!]
{\caption{Parameters of the flux tubes for the cases considered}}
\vspace{0.2cm}
{\centerline{
\begin{tabular}
{|c|c|c|c|} \hline \hline
 & AGB & RGB - 1 & RGB - 2 \\
\hline
  \hline
\noindent
$a_0$ (km) &  65 & 65 & 150 \\
  \hline
\noindent
$\hat v$ (km/sec)& 1.45 & 0.098 & 0.018 \\
  \hline
\noindent
$v_0$ (km/sec) & 6.80 & 0.336 & 0.062 \\
  \hline
  \noindent
 $B_0$ (G) & 5$\times 10^6$ & 3.8$\times 10^5$ & 4.8$\times 10^4$ \\
  \hline
 \noindent
 $B_{E}$ (G) & 8600  & 3450 & 440 \\
  \hline
  \hline
\end{tabular}}}
\end{table}
%\clearpage

\section{Conclusions}

We have investigated the consequences resulting from the assumption
that buoyant magnetic flux tubes may be responsible for the
transport of matter processed in the neighborhood of the hydrogen
burning shell. Such an assumption is not unrealistic, not only on
the basis of the solar scenario, but also because the different
characteristics of stellar dynamo processes have been recently
suggested, on observational grounds, to be linked to the
establishment of deep mixing (B\"ohm-Vitense 2007).

It is assumed that both RGB and AGB stars have dynamos producing
toroidal magnetic flux tubes in that region. From the observational
requirements that matter has undergone nuclear processing and has
been transported through the radiative zone to the convective
envelope, we know the mass flow rates ($\dot M$) that are needed and
the temperature of processing ($T_P$). Furthermore, the structure
and evolutionary time scales for both the RGB and AGB stars are
known from well established stellar models. Since evidence for
chemical peculiarities induced by CBP is shown only by relatively
low mass red giants (below 2 $-$ 3 \ms), our calculations were done
for a star of initial mass 1.5 \ms, with half solar metallicity.
Assuming transport by buoyant magnetic flux tubes, it was possible
to calculate the magnetic fields $B_0$ required at the processing
zone to give the necessary mass transfer rates. It is shown that to
supply the processed material to the AGB star requires large
magnetic fields near the H shell, at the level $B_0 \sim 5\times
10^6$ G at the bottom of the CBP region and $B_{E} \sim 9\times
10^3$ G at the base of the convective envelope. For the RGB case, it
was found that the required fields are much smaller, with $B_0$ in
the range 5$\times$10$^4$ to 4$\times$10$^5$ G and with $B_{E}$ in
the range 450$-$3500 G, depending on how the flux tube size was
estimated. This processing and transport will produce
$^{12}$C/$^{13}$C ratios of 10 to 12 in the envelope at the end of
the RGB evolution and would extensively destroy Li.

It follows that, if magnetic buoyancy is the means of transport for CBP, then very high
fields are required at great depths for the AGB phase and substantial, but much more modest,
fields are required for the RGB phase.

There is no direct observational evidence of strong magnetic fields
in either RGB or AGB stars. However, it is quite plausible that the
rather modest fields we find at the base of the envelope would be
greatly subdued when engulfed in the very massive overlying
convecting stellar envelope. A rough estimate for AGB stars, using
flux conservation, gives surface fields of $\lesssim 20 $G, in line
with observational estimates by Herpin et al. (2006). This is
compatible with the fact that AGB stars are not observed to have
X-ray emitting coronae. For the RGB case the surface fields would
again be small, in the range from $\lesssim$ 1 to a few G.

With regard to the high fields found to be necessary for the AGB case, we note that
plasma jets of magnetic origin have been observed for some planetary nebulae
(cf. Kastner et al. 2003). A planetary nebula is the result of the end of AGB
evolution, when the envelope is blown off, leaving a white dwarf remnant. Blackman
et al. (2001) considered the development of dynamos in AGB stars as the origin of
magnetic fields shaping planetary nebulae. In their calculations they inferred fields
of $\sim$ 5$\times$10$^4$ G at the base of the AGB convective envelope (at a radius
of $\sim$ 1.5 $R_{\odot}$) in order to explain the collimation. $B$ values of up to
a few $\times$10$^6$ G were inferred in case of low filling factors (which is in
fact our case, due to the low values of $f_m$). There is thus
independent evidence for large internal magnetic fields  in AGB stars.

It is further of note that white dwarfs, which are the end product
of AGB evolution, do sometimes show very high fields. Recent studies
have revived the hypothesis that these fields might be fossil
remnants of the stellar ones (Tout et al. 2004; Wickramasinghe \&
Ferrario 2005; Ferrario \& Wickramasinghe 2005; Valyavin et al.
2006). Of special importance for this hypothesis is the
identification of magnetic configurations, with toroidal components,
that can remain stable for extremely long periods (Braithwaite and
Spruit 2006). Many white dwarfs either do not show magnetic fields,
or show them at the kG level; nevertheless fields larger than 1 MG,
and up to 2 Gigagauss, are present in 10$-$20 \% of the available
white dwarf sample (Liebert et al. 2003; Schmidt et al. 2003; Jordan
et al. 2007). Most super-magnetic objects seem to be of relatively
high mass ($\sim$0.9 \ms), but systems with $B = 1 - 100 \times
10^6$ G and with masses $M \le 0.75$ \msb are not infrequent (cf.
Liebert et al. 2003, Tables 1-3). These observations suggest that
the high fields of such white dwarfs might be inherited from the AGB
precursor, with high enhancement factors ensuing upon the expulsion
of the envelope.

From the arguments presented in this report we conclude that
magnetic buoyancy is a very plausible mechanism for transporting
material from near the H shell into the convective envelope, for
both RGB and AGB stars. This requires very high internal fields for
AGB stars with extensive CBP. In our presentation we have only given
phenomenological arguments regarding the transport, assuming that
the stars provide the required magnetic fields. It is our hope that
the broader considerations presented here may stimulate intensive
MHD modeling of low mass stars with very extended envelopes.

\acknowledgements 
The authors would like to thank Hans Hornung for
views on the possible values of $C_D$. M.B and A.C acknowledge
support from MURST, under contract PRIN2006-022731. G.J.W.
acknowledges the support of DOE-FG03-88ER13851 and the generosity of
the Epsilon foundation. CALTECH contribution \#9178(1123). KMN is
supported by the US Department of Energy, Office of Nuclear Physics,
under contract No. DE-AC02-06CH11357.  The authors would like to
thank the reviewer (clearly an expert of MHD) for a most
constructive and useful critical review.  We would like to dedicate
this paper to Gene Parker and Bob Leighton for their pioneering
Solar activities.

 \end{document}